\documentclass[fleqn,usenatbib]{mnras}

\usepackage[T1]{fontenc}
\usepackage{ae,aecompl}
\usepackage{times}
\usepackage{color}
\usepackage{soul}

\usepackage{graphicx}	% Including figure files
\usepackage{amsmath}	% Advanced maths commands
\usepackage{amssymb}	% Extra maths symbols

\newcommand{\cf}{cf.,~}
\newcommand{\ie}{i.e.,~}
\newcommand{\eg}{e.g.,~}

\title[Stability and maximum mass of differentially rotating stars]{On the
  stability and maximum mass of differentially rotating relativistic
  stars}

\author[Weih, Most and Rezzolla]
{
Lukas R. Weih$^{1}$ \thanks{E-mail: weih@th.physik.uni-frankfurt.de},
Elias R. Most$^{1}$
and Luciano Rezzolla$^{1,2}$
\\
$^{1}$Institut f\"ur Theoretische Physik, Goethe Universit\"at Frankfurt,
Max-von-Laue-Str.1, 60438 Frankfurt am Main, Germany\\
$^{2}$Frankfurt Institute for Advanced Studies, Ruth-Moufang-Str. 1, 60438
Frankfurt am Main, Germany 
}

\pubyear{2017}

\begin{document}
%\label{firstpage}
\maketitle

\begin{abstract}
The stability properties of rotating relativistic stars against prompt
gravitational collapse to a black hole are rather well understood for
uniformly rotating models. This is not the case for differentially
rotating neutron stars, which are expected to be produced in catastrophic
events such as the merger of binary system of neutron stars or the
collapse of a massive stellar core. We consider sequences of differentially
rotating equilibrium models using the $j$-constant law and by combining them
with their dynamical evolution, we show that a sufficient
stability criterion for differentially rotating neutron stars exists
similar to the one of their uniformly rotating counterparts. Namely:
along a sequence of constant angular momentum, a dynamical instability
sets in for central rest-mass densities slightly below the one of the
equilibrium solution at the turning point. In addition, following
\citet{Breu2016}, we show that ``quasi-universal'' relations can be found
when calculating the turning-point mass. In turn, this allows us to
compute the maximum mass allowed by differential rotation, $M_{\rm
  max,dr}$, in terms of the maximum mass of the nonrotating
configuration, $M_{_{\rm TOV}}$, finding that $M_{\rm max, dr} \simeq
\left(1.54 \pm 0.05\right) M_{_{\rm TOV}}$ for all the equations of state
we have considered.
\end{abstract}


%%%%%%%%%%%%%%%%%%%%%%%%%%%%%%%%%%%%%%%%%%%%%%%%%%

%%%%%%%%%%%%%%%%% BODY OF PAPER %%%%%%%%%%%%%%%%%%

%-----------------------------------------------------------------------
\section{Introduction}
\label{sec:intro}
%-----------------------------------------------------------------------

%
Simulations of binary neutron-star mergers have shown that the merger
remnant is a differentially rotating neutron star [see
  \citet{Baiotti2016} for a recent review]. Depending on the total mass,
mass ratio and equation of state (EOS) of the binary, the remnant may be
a hypermassive neutron star (HMNS) and collapse to a black hole on a
timescale $\lesssim 1\,{\rm s}$, or be a long-lived supramassive neutron
star, or even a stable neutron star. The possible outcome has important
implications on the gravitational-wave signal of such a merger, whose
detection could be imminent. While the simulations of merging binaries
are computationally expensive, an equilibrium solution of a
differentially rotating neutron star is a good and inexpensive
approximation for certain types of problems. Thus, a lot of studies have
been dedicated to such solutions, either for their equilibrium
\citep{Baumgarte00b, Lyford2003, Ansorg2009, Gondek2016, Studzinska2016}
or for their dynamics \citep{Duez:2006qe, Giacomazzo2011}. Following
these studies, we here use the ``$j$-constant'' law to model 
differential rotation (see also Sec. \ref{sec:id}); while commonly used,
this rotation law is not a good approximation for
the differentially rotating object produced in binary neutron-star
mergers [see, \eg \citet{Kastaun2014, Hanauske2016}], but may be considered
as a first step towards studying the stability properties of differentially
rotating neutron stars.

The stability of neutron stars is a classical problem in general
relativity and one of its most important results is the so-called
``turning-point criterion'' by \citet{Friedman88}. It states that along a
sequence of nonrotating relativistic stars, secular instability sets in at
the maximum of this sequence, \ie at the turning point. On the other
hand, \citet{Takami:2011} found that for uniformly rotating stars this is
just a sufficient, not a necessary criterion. The onset of dynamical
instability is instead marked by the neutral-stability line, \ie where
the eigenfrequency of the fundamental mode of oscillation vanishes. The
neutral-stability line and the turning-point line coincide for
nonrotating neutron stars, but their difference grows with increasing
angular momentum \citep{Takami:2011}.

We here analyse the stability of differentially rotating neutron stars by
carefully choosing 53 equilibrium models along sequences of constant
angular momentum close to their respective turning points and evolving
them dynamically in full general relativity. The choice of models close
to their turning points is based on the conjecture that a
neutral-stability line as found in \citet{Takami:2011} also exists for
differentially rotating neutron stars. Arguing that the neutral-stability
line is reasonably close to the turning points, the latter are routinely
used to find the stability limit not just for uniformly
\citep{Baiotti04}, but also for differentially rotating neutron stars
\citep{Bauswein2017, Kaplan2014}. Using fully general-relativistic
numerical simulations we show that this assumption is indeed correct and
the turning-point line can be used as an approximation for finding the
threshold mass for prompt collapse to a black hole. Our results apply in
particular to the class of solutions of ``type A'', as classified by
\citet{Ansorg2009}. These differentially rotating stars always have the
maximum of the rest-mass density at the stellar center, are spheroidal,
and possess a mass-shedding limit. When considering the properties of the
remnant of binary neutron-star merger simulations, this class of
solutions appears to be the most realistic one.

In addition, we present a ``quasi-universal'' relation that allows us to
determine the turning-point mass of differentially rotating neutron stars
independently of the underlying EOS. This relation is an extension to
what was found by \citet{Breu2016} for uniformly rotating stars and is in
agreement with what was recently presented by \citet{Bozzola2017}. More
importantly, using this relation, we can estimate the maximum mass
allowed by differential rotation, $M_{\rm max, \, dr}$, in terms of the
maximum mass of the nonrotating configuration, $M_{_{\rm TOV}}$. We find
that $M_{\rm max,dr} \simeq \left(1.54 \pm 0.05\right) M_{_{\rm TOV}}$
for all the EOSs considered.

The organization of the paper is as follows. In Sec. \ref{sec:id} we
discuss the equilibrium models, while the numerical setup of the
numerical evolution is described in Sec. \ref{sec:numset}, followed by
the presentation of the results in Sec. \ref{sec:results}. The
quasi-universal relation for the maximum mass is discussed in
Sec. \ref{sec:unirel}, while our results are summarised in
Sec. \ref{sec:conc}. Unless stated differently, we use units in which
$c=G=M_\odot=1$ and call simply ``mass'' the gravitational mass.

%-----------------------------------------------------------------------
\section{Numerical setup and Initial data}
\label{sec:nsaid}
%-----------------------------------------------------------------------
\subsection{Initial Data}
\label{sec:id}

\begin{figure}
     \centering \includegraphics[width=0.75\columnwidth]{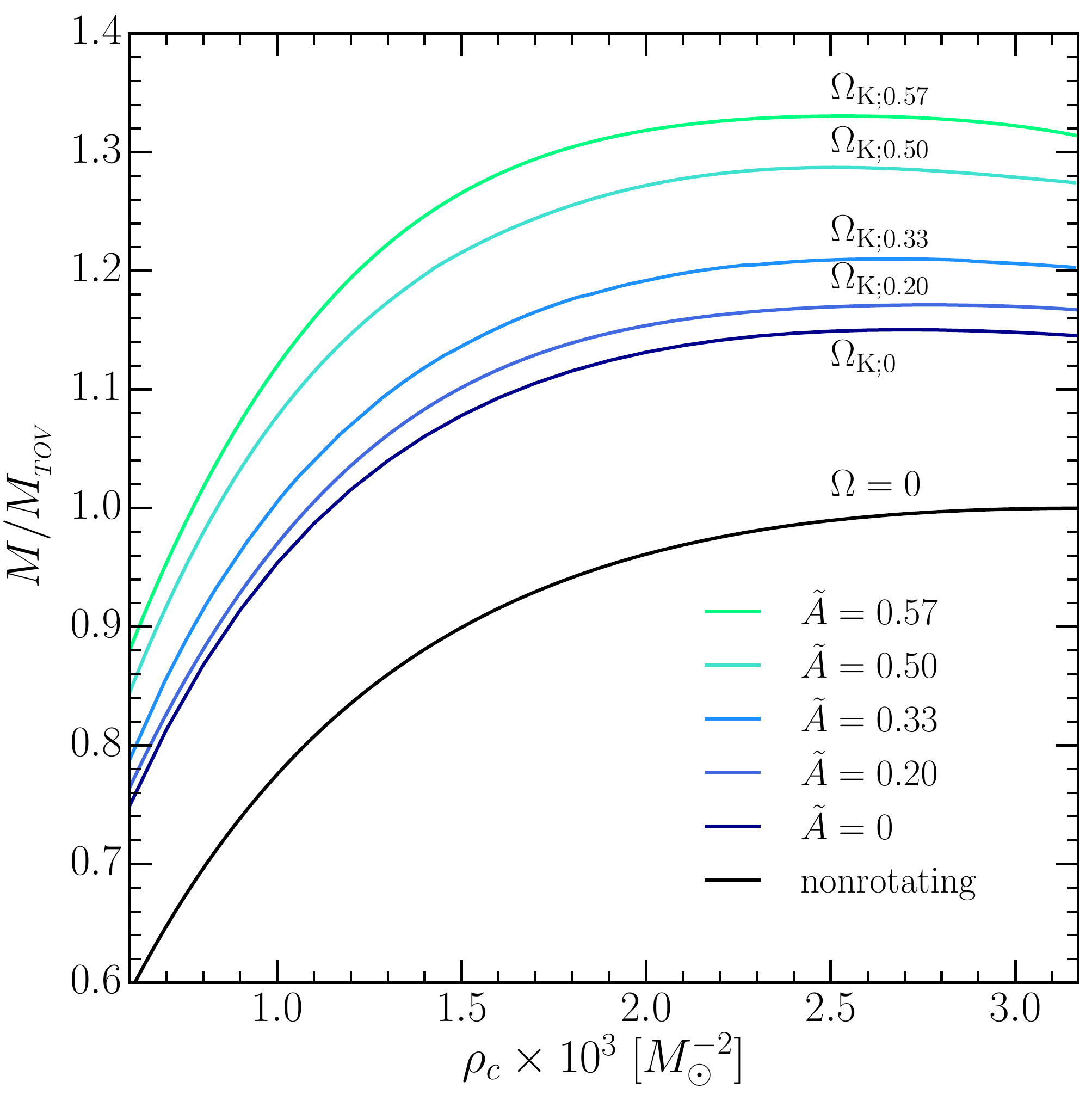}
     \caption{Generic sequences of equilibrium solutions with polytropic
       EOS: nonrotating
       models ($\Omega=0$), mass-shedding limits of uniformly rotating
       models ($\Omega_{\rm K;0}$), and mass-shedding limits of
       differentially rotating models with different degrees of
       differential rotation ($\Omega_{\rm K;0.20} - \Omega_{\rm
         K;0.57}$).}
     \label{fig:zero}
\end{figure}

The initial data represents axisymmetric models of self-gravitating
matter configurations in equilibrium and is computed numerically making
use of the \texttt{RNS} code \citep{Stergioulas95}, which solves the
Einstein equations together with the equation of hydro-stationary
equilibrium. Differential rotation is then introduced via a rotation-law
function $F(\Omega)$ that depends on the angular velocity $\Omega$ only
and determines the star's rotation profile. For our equilibrium models we
chose the commonly used $j$-constant law
\begin{equation}
\label{eq:rotlaw}
F(\Omega)=A^2(\Omega_c-\Omega)\,,
\end{equation}
where $\Omega_c$ is the central angular velocity and $A$ is a free
parameter with dimension length that determines the degree of
differential rotation. We adopt the parametrization $\tilde{A}:=r_e/A$
with $r_e$ being the equatorial coordinate radius\footnote{Another
  commonly used parametrization is $\hat{A}:=A/r_e=\tilde{A}^{-1}$.}. In
this way, the limit of uniform rotation is obtained for $\tilde{A}=0$,
while higher values of $\tilde{A}$ lead to higher degrees of differential
rotation, \ie steeper rotation profiles. Another consequence of
differential rotation is that for higher values of $\tilde{A}$, higher
angular momenta $J$ can be reached, which, in turn, shifts the
mass-shedding limit to higher masses.  This is shown in
Fig. \ref{fig:zero}, which reports sequences of nonrotating models
($\Omega=0$), together with the mass-shedding limits of uniformly
rotating models ($\Omega_{\rm K;0}$), and of differentially rotating
models with different degrees of differential rotation ($\Omega_{\rm
  K;0.20}-\Omega_{\rm K;0.57}$).

\begin{figure*}
     \centering
     \includegraphics[width=0.75\columnwidth]{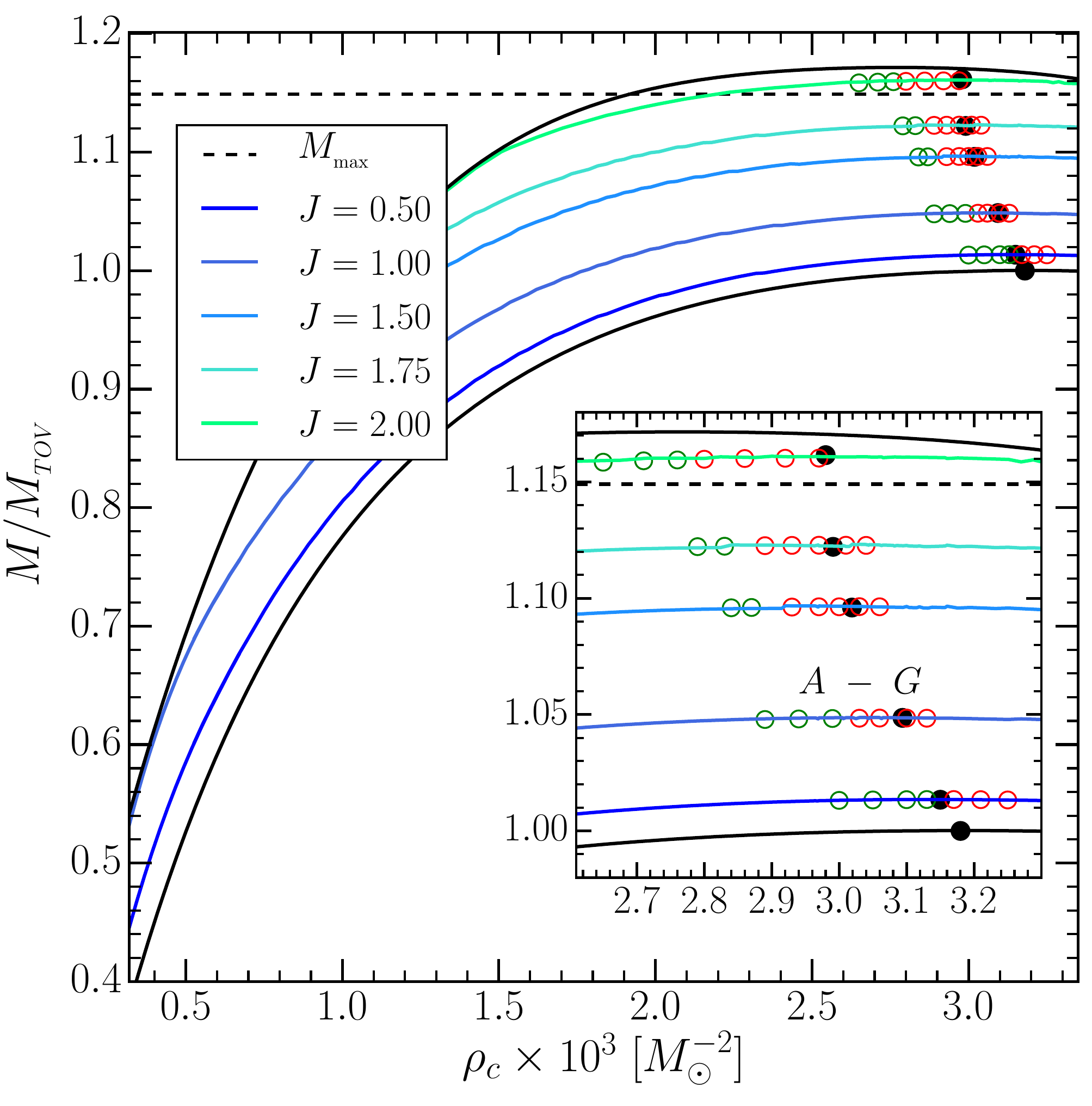}
     \hspace{2.5cm}
     \includegraphics[width=0.75\columnwidth]{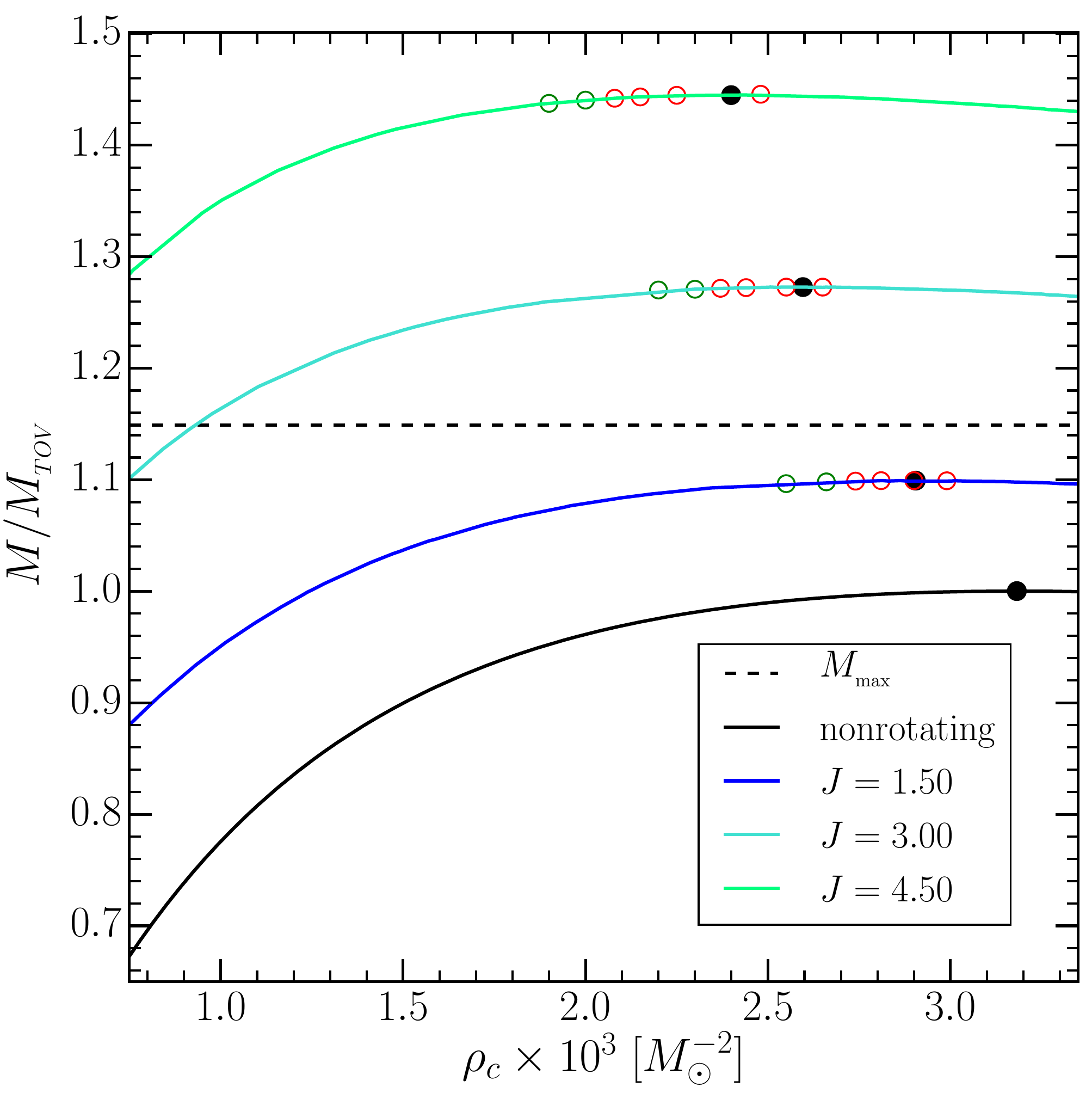}
     \caption{\textit{Left panel:} Sequences of constant angular momentum
       for $\tilde{A}=0.2$. The lower and upper black line refer to the
       nonrotating and mass-shedding sequences, \ie $\Omega=0$ and
       $\Omega_{\rm K;0.20}$, respectively. Black filled circles denote
       the turning points, while the open circles mark the dynamically
       evolved models. The red circles, collapse on a dynamical
       timescale, while the green ones do not. Models $A$-$G$ are
       labelled from left to right, while an horizontal dashed line marks
       the maximum mass of uniformly rotating models, $M_{\rm
         max}$. \textit{Right panel:} Same as in the left panel, but for
       $\tilde{A}=0.77$. Note that an overall mass shedding limit does
       not exists for this degree of differential rotation.  }
     \label{fig:equilmodels1}
\end{figure*}

The initial data for the dynamical evolution is modeled with a polytropic
EOS with $\Gamma=2.0$ and $K=100$ \citep{Rezzolla_book:2013} and we
compute a wide range of models by specifying the values for $\tilde{A}$,
the central rest-mass density $\rho_c$, and the ratio of polar to
equatorial radius. We then arrange these models as sequences of constant
angular momentum\footnote{The variance of the angular momentum from a
  constant value is $\lesssim 1\%$.} parametrized by $\rho_c$. Such
sequences are shown in Fig. \ref{fig:equilmodels1}, which also reports
the location of the initial data used for the dynamical
evolution with open circles.
On the other hand, for the quasi-universal relation presented
in Sec. \ref{sec:unirel} the equilibria are computed in the same way, but
with eight nuclear-physics cold EOSs in tabulated form: DD2, NL3, SLy,
SKa, SK272, SK255, SFHX, and SFHO [see \citet{Baiotti2016} for a list of
  references to these EOSs which we omit for compactness here]. Note that we
have fixed the electron fraction assuming beta equilibrium and the temperature to
be the lowest table entry for hot EOS. We point out that all the EOS used here
are compatible with the $2M_\odot$ constraint of nonrotating
neutron stars \citep{Demorest2010, Antoniadis2013}.

%------------------------------------------------------------------------------

\subsection{Numerical Setup}
\label{sec:numset}

From the initial data the necessary quantities of the $3+1$ formalism are
computed and mapped on a Cartesian grid. In order to extend the grid and
increase the resolution within the star fixed mesh refinement is used,
which is provided by the \texttt{Carpet} driver
\citep{Schnetter-etal-03b}. Two refinement levels are used with an inner
grid resolution of $h_1=0.1\,M_\odot$ ($\sim 148\,{\rm m}$), and the
outer grid with half that resolution. The results have been confirmed to
be the qualitatively same when employing a higher resolution of
$h_1=0.07\,M_\odot$ ($\sim 100\,{\rm m}$) and a lower resolution of
$h_1=0.15\,M_\odot$ ($\sim 220\,{\rm m}$).  The boundary of the inner
level is placed at $12\,M_\odot$ ($\sim 17.7\,{\rm km}$) so that the star
lies in all cases well within this inner region. The outer boundary is
placed at twice that distance (it has been verified that the outer
boundary at four times the inner one yields the same results). In order
to save computation time the neutron star's symmetry is exploited and a
reflection symmetry across the $z=0$ plane is adopted. The spacetime is
evolved using the fourth-order finite-differencing code
\texttt{McLachlan} \citep{loeffler_2011_et} and apparent horizons are
computed by \texttt{AHFinderDirect} \citep{Thornburg2003:AH-finding}. The
\texttt{McLachlan} code solves the equations of the CCZ4 formulation
\citep{Alic:2011a} using the $1+\log$ slicing condition for the lapse
\citep{Alcubierre99d} and the Gamma-driver condition for the shift
\citep{Alcubierre01a}. On the other hand, the evolution of the
hydrodynamic quantities is done by the high-order \texttt{WhiskyTHC} code
\citep{Radice2012a, Radice2013b, Radice2013c} employing an ideal-fluid
EOS with $\gamma=2$. It employs a fifth-order monotonicity preserving
flux reconstruction (MP5) \citep{suresh_1997_amp} and is coupled to a
third-order Runge-Kutta time integrator with a Courant factor of $0.15$.

In principle, the truncation error of the initial data is sufficient to
excite perturbations in the star and hence either quasi-normal mode
oscillations or trigger the gravitational collapse
\citep{Baiotti04}. However, given an unstable equilibrium model, the
properties of our truncation error always lead the star to migrate to the
stable branch of solutions rather than collapsing to a black hole
\citep{Font02c}. Because for models close to the turning point the
solution on the unstable branch is very close in mass to the stable one,
such a migration is almost indistinguishable from the actual oscillations
of a stable star. A much clearer distinction is possible if the unstable
models collapse, which can be ensured by decreasing the initial angular
velocity according to
\begin{equation}
\label{eq:pert1}
\delta \Omega(r,\theta,\phi) = \mathcal{A}\,\Omega(r,\theta,\phi)\,,
\end{equation}
where the amplitude $\mathcal{A}$ of that relative perturbation has to be
chosen carefully: a perturbation that is too strong might push a
perfectly stable configuration over the stability limit and force it to
collapse. To prevent this, the amplitude was gauged using results known
for uniformly rotating stars and an analytic expression for the
neutral-stability line for uniformly rotating neutron stars can be
obtained by setting to zero Eq. (3) of \citet{Takami:2011}, \ie
\begin{equation}
\label{eq:takami}
\beta(\rho_c) = -\left({\sum_{n=0}^{n=5} a_n \rho_c^n }\right) /\left({
  \sum_{n=0}^{n=5} b_n \rho_c^n}\right)\,,
\end{equation}
where $\beta$ is the ratio of kinetic to gravitational binding energy and
the parameters $a_n$ and $b_n$ are given by \citet{Takami:2011}. Next,
using the \texttt{RNS} code, we compute a sequence of uniformly rotating
stars with constant angular momentum $J=1.0$. The intersection of
$\beta(\rho_c)$ for this sequence with Eq. \eqref{eq:takami} yields the
location $\rho_c^{\rm crit}$ of the stability limit for $J=1.0$. We then
select models of our sequence with central rest-mass densities that are
slightly smaller and slightly larger than $\rho_c^{\rm crit}$ and set
$\mathcal{A}$ as the value for which the latter configuration
collapses while the former does not. This value is then used to perturb
the differentially rotating neutron stars.

%-----------------------------------------------------------------------
\section{Numerical results and Analysis}
\label{sec:results}
%-----------------------------------------------------------------------

\begin{figure}
     \centering
     \includegraphics[width=0.75\columnwidth]{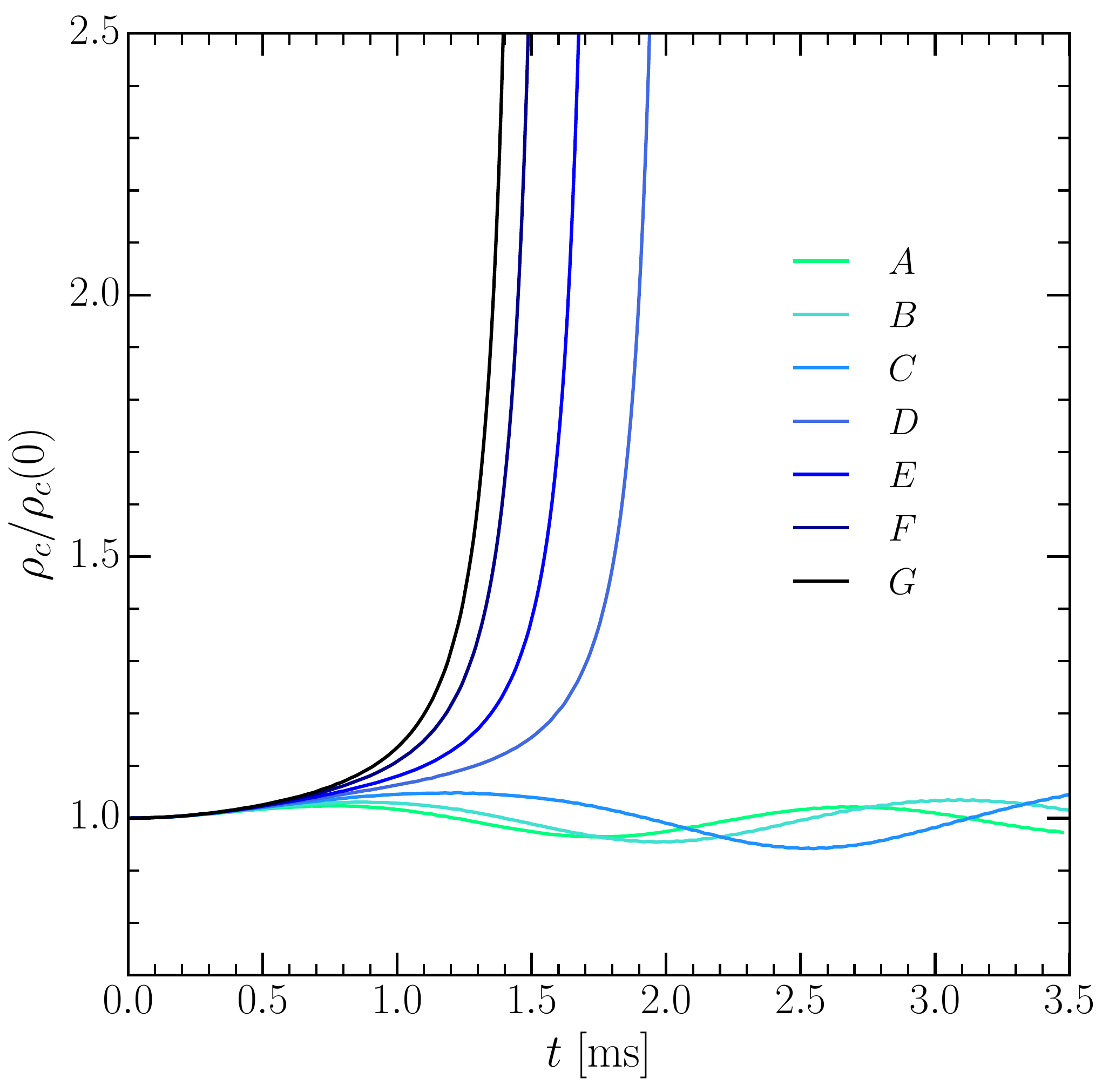}
     \caption{Evolution of the central rest-mass density normalised to
       its initial value. All models lie on the same sequence of constant
       angular momentum $J=1.0$ and have a degree of differential
       rotation of $\tilde{A}=0.2$ (\cf Fig. \ref{fig:equilmodels1}).}
     \label{fig:timeevol}
\end{figure}

In Fig. \ref{fig:timeevol} the time evolution of the central rest-mass
density $\rho_c$ is shown for several models with $J=1.0$ and
$\tilde{A}=0.2$ around the turning point of this sequence. The diverging
curves clearly show the prompt collapse to a black hole and this is further
confirmed by the detection of an apparent horizon. A collapsing model
does so after $\sim 1-2 \rm ms$ and shows the same dynamics as was
already observed for differentially rotating neutron stars by
\citet{Giacomazzo2011}. In contrast, the three models with the lowest
densities do not collapse and show only small oscillations induced by the
initial perturbation, which have been verified to vanish for unperturbed
initial data and to correspond to the fundamental mode of oscillation
(the frequency decreases near the threshold to stability).

Whether a model collapses or not should be independent of the kind of
perturbation that is applied. Therefore, we additionally tested two
models, \ie one that was found to be stable and one unstable using the
previous perturbation. We then evolved these models applying an
inwards-directed radial perturbation of the form
\begin{equation}
\label{eq:pert2}
\delta v_r(r,\theta,\phi) = \mathcal{B} \, |v(r,\theta,\phi)|\,,
\end{equation}
where the absolute velocity is calculated as $|v|=\sqrt{g_{ij}v^iv^j}$
and the amplitude $\mathcal{B}$ is gauged in the same way as was done for
$\mathcal{A}$ in \eqref{eq:pert1}. Although the collapse dynamics differs
slightly depending on the perturbation, the fact whether a model
collapses or not does not.

What can be seen in Fig. \ref{fig:timeevol} looks qualitatively the same
for all other sequences, \ie the high-density models collapse while the
low-density ones do not. Referring to Fig. \ref{fig:equilmodels1}, the
stability properties of all the evolved models with $\tilde{A}=0.20$ can
be easily seen: black filled circles mark the turning points of the
sequences of constant angular momentum, while red circles mark
configurations that were found to collapse, and green circles mark the
stable ones. The same analysis has been repeated for 18
models with a higher degree of differential rotation, \ie
$\tilde{A}=0.77$. For this value no mass-shedding limit exists, because
type-A stars cannot be found for higher $J$. The overall picture looks
qualitatively similar to the one shown in Fig. \ref{fig:equilmodels1},
although the stability limit shifts even further to the low-density side
of the turning point when $J$ is increased.

Clearly, all the differentially rotating neutron stars to the
high-density side of their respective turning point are unstable. As the
turning point shifts to lower densities with increasing angular momentum,
we can thus confirm what conjectured by \citet{Kaplan2014}: All the
differentially rotating stars with central densities at or higher than
the critical central density of a nonrotating model are unstable to
gravitational collapse. While they conjectured this for all HMNSs, our
results show it is holds for all differentially rotating neutron stars of
type A, including the non-hypermassive ones.

We should also note that the configurations on the low-density side of
their respective turning point are not unconditionally stable. As is the
case for uniformly rotating stars, the actual neutral-stability line is
on the left of the turning-point line, so that stellar models that are on
the left of the turning-point but on the right of the neutral-stability line 
may actually also be unstable [see \citet{Takami:2011} for a discussion]. With 
increasing angular momentum the stability limit shifts to the low-density side 
of the turning point, but it is still reasonably close to the turning point, 
so that the approximation of taking the turning point as the stability limit
is valid, at least for small values of $J$.

The models in Fig. \ref{fig:equilmodels1} above the horizontal dashed
line are HMNSs but have the same overall behavior as the non-hypermassive
ones. It appears, therefore that the assumption that all HMNSs are
dynamically unstable is not justified. However, HMNSs might still be
secularly unstable and pass the neutral-stability line after
redistributing their angular momentum.

%------------------------------------------------------------------------------
\section{Maximum Mass and Universal Relations}
\label{sec:unirel}
%------------------------------------------------------------------------------
%
\begin{figure}
     \centering \includegraphics[width=0.75\columnwidth]{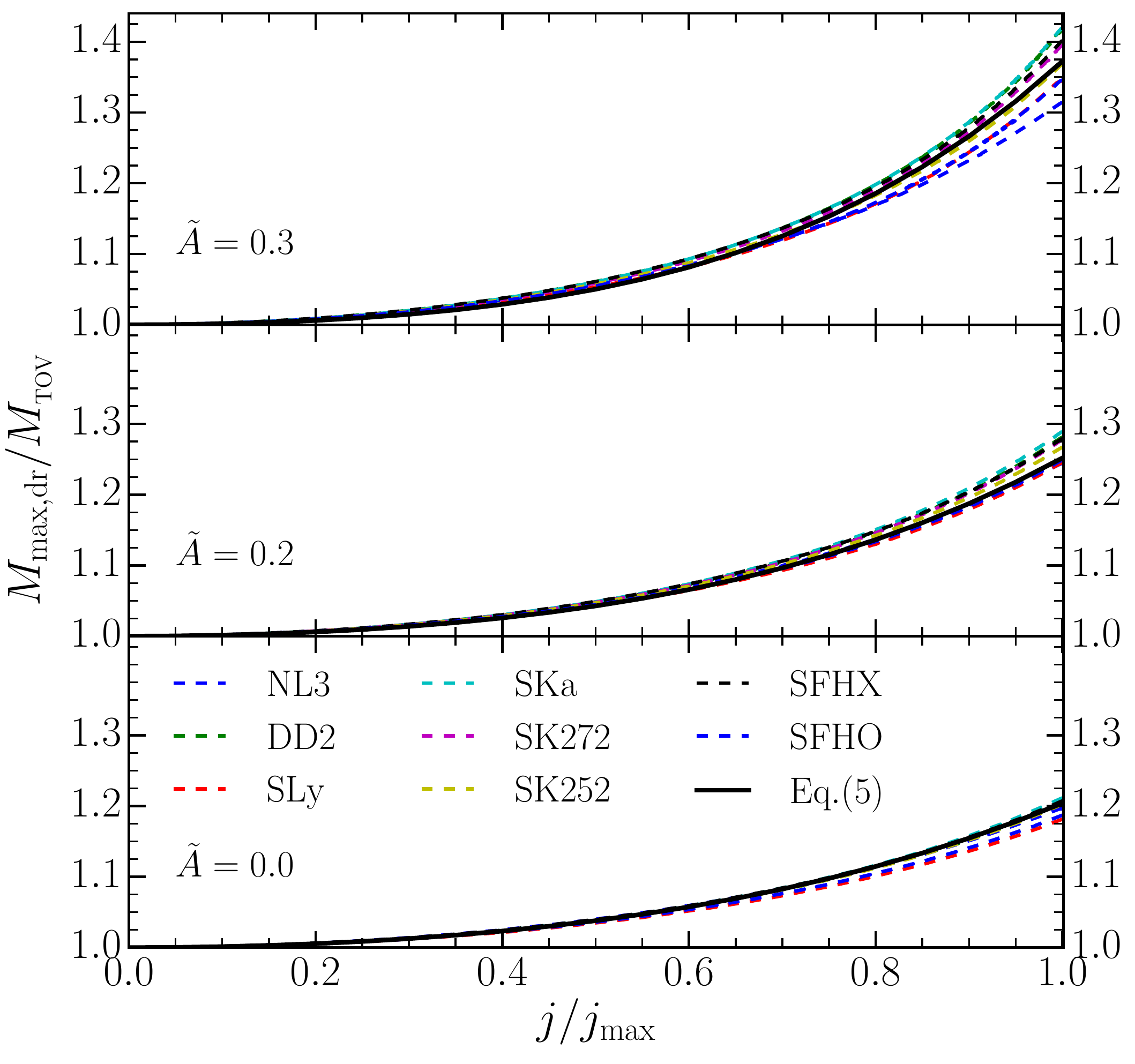}
     \caption{Normalised turning-point mass over normalised dimensionless
       angular momentum for eight different EOSs. The three panels
       correspond to three different degrees of differential rotation and
       the best-fitting functions are shown as thick solid black lines.}
     \label{fig:uni1}
\end{figure}

\begin{figure}
     \centering
     \includegraphics[width=0.75\columnwidth]{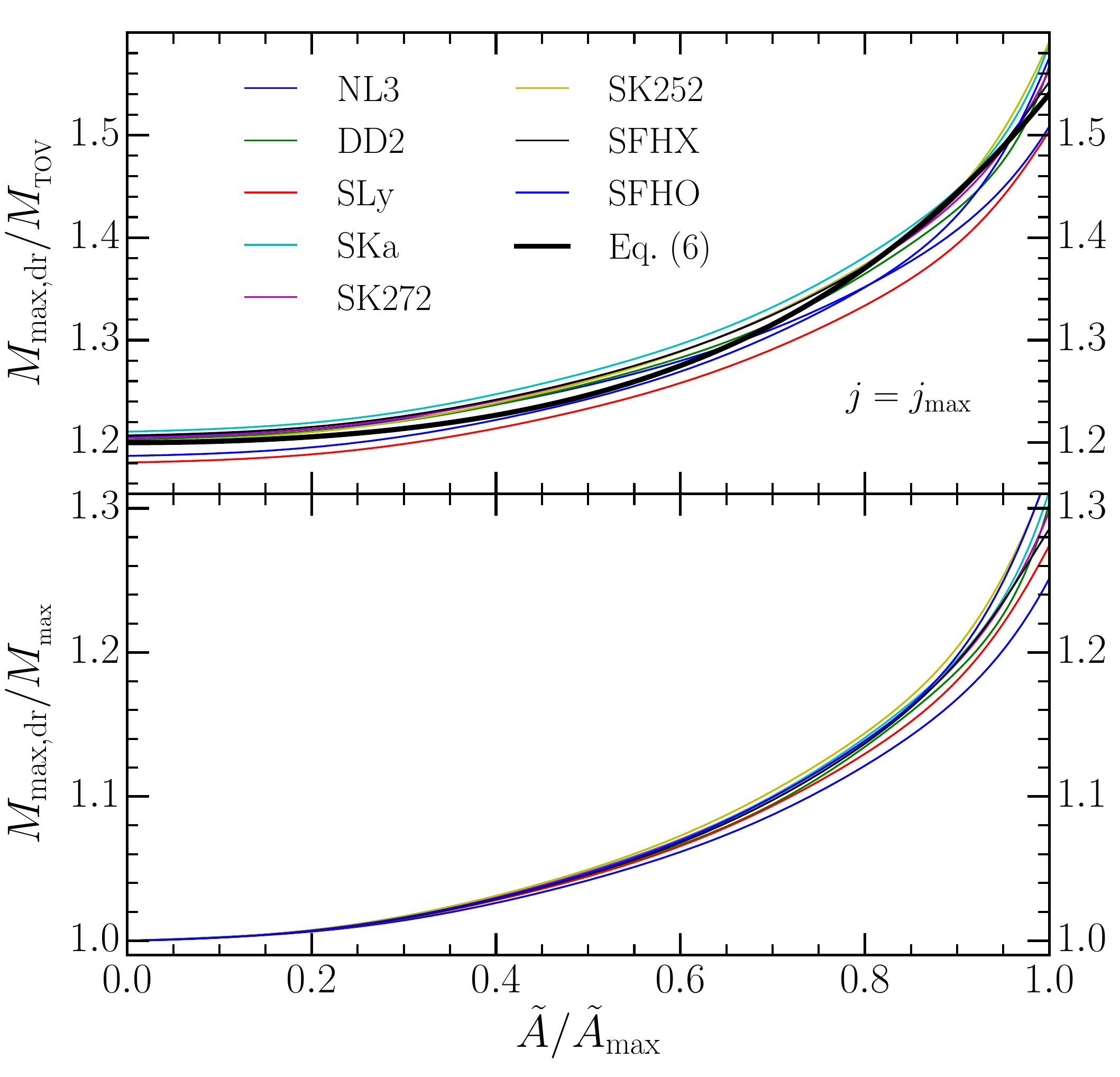}
     \caption{Maximum mass as a function of the normalised degree of
       differential rotation. In the upper panel the data is compared to
       the fit \eqref{eq:maxmass}, while the lower panel further
       highlights the universal behavior by normalising the mass to the
       highest value obtained for uniform rotation, $M_{\rm max}$.}
     \label{fig:uni2}
\end{figure}

Figure \ref{fig:uni1} shows the normalised turning-point mass $M_{\rm
  max,dr}$ as a function of the normalised dimensionless angular momentum
$j/j_{\rm max}$, for all of the eight cold tabulated EOSs
considered. Here, $j:=J/M^2$ and $j_{\rm max}$ is instead the highest
specific angular momentum which is at the mass-shedding limit in the case
of uniform rotation, but not for $\tilde{A}>0$. Lines of the same type
are collected in three bundles -- each consisting of eight coloured
curves (one for each EOS) -- corresponding to three representative values
of $\tilde{A}$. The bundle in the lower panel corresponds to $\tilde{A}=0$, 
hence to uniform rotation and should therefore be compared with the right
panel of Fig. 1 in \citet{Breu2016}. Figure \ref{fig:uni1} clearly highlights 
that the turning-point mass increases with growing $j/j_{\rm max}$ in a
quasi-universal manner, \ie in a way that is almost insensitive to the
EOS. As discussed by \citet{Breu2016}, in the case of uniform rotation
(\ie $\tilde{A}=0$), the turning-point mass reaches a maximum at $M_{\rm
  max} \simeq 1.2\,M_{_{\rm TOV}}$. Our results show that this general
behaviour remains the same also for non-zero values of $\tilde{A}$, as is
evident from the two upper panels of Fig. \ref{fig:uni1} that refer to
differentially rotating models. As done by \citet{Breu2016} we can now 
obtain corresponding fits, reported with thick lines of the same type and
have the form
\begin{equation}
\label{eq:fit}
\frac{M_{\rm max,dr}(j,\tilde{A})}{M_{_{\rm TOV}}} = 1+
a_1(\tilde{A})\left(\frac{j}{j_{\rm max}} \right)^2 +
a_2(\tilde{A})\left( \frac{j}{j_{\rm max}} \right)^4\,.
\end{equation}
Because $M_{\rm max,dr}$ is a function of both $j$ and $\tilde{A}$, the
fitting parameters $a_1$ and $a_2$ are not constant, as in
\citet{Breu2016}, but depend now on the degree of differential rotation
via $\tilde{A}$, with $a_1=0.13,0.14,0.14$ and $a_2=0.07,0.11,0.23$ for
$\tilde{A}=0,0.2,0.3$, respectively, and with relative variances that are
$\lesssim 10\%$. We can collect the maximum mass $M_{\rm max,
  dr}(\tilde{A})$ found at $j/j_{\rm max}=1$ for all of the eight EOSs
and study its behaviour when $\tilde{A}$ varies in the range $[0,
  \tilde{A}_{\rm max}]$, where $\tilde{A}_{\rm max}$ is the maximum
degree of differential rotation for which a mass-shedding limit can still
be found, and which obviously depends on the EOS\footnote{We note that
  stars of ``type C'' in the classification of \citet{Ansorg2009} do not
  posses a mass-shedding limit and hence a value for $\tilde{A}_{\rm
    max}$.}. Using this procedure we are then able to concentrate on
models that have the largest possible specific angular momentum and study
how the maximum mass changes as function of $\tilde{A}$. Interestingly,
we find a quasi-universal behaviour also in this case, which we model as
\begin{equation}
\label{eq:maxmass}
\frac{M_{\rm max,dr}(j_{\rm max},\tilde{A})}{M_{_{\rm TOV}}} = 1.2+
b_1\left(\frac{\tilde{A}}{\tilde{A}_{\rm max}}\right)^2+
b_2\left(\frac{\tilde{A}}{\tilde{A}_{\rm max}}\right)^4\,,
\end{equation}
where $b_1=0.135$ and $b_2=0.206$. The global behaviour of the fitting
function \eqref{eq:fit} for the quantity $M_{\rm max,dr}(\tilde{A})$ is
shown in Fig. \ref{fig:uni2}. Using \eqref{eq:maxmass} for
$\tilde{A}=\tilde{A}_{\rm max}$, we then find ``absolute'' maximum mass
(\ie the maximum of the maxima) for a star in differential rotation of
type A to be $M_{\rm max, \, dr} \simeq \left(1.54 \pm 0.05\right)
M_{_{\rm TOV}}$, where the error estimate results from considering the
largest error in our fits.

\begin{figure}
     \centering
     \includegraphics[width=0.65\columnwidth]{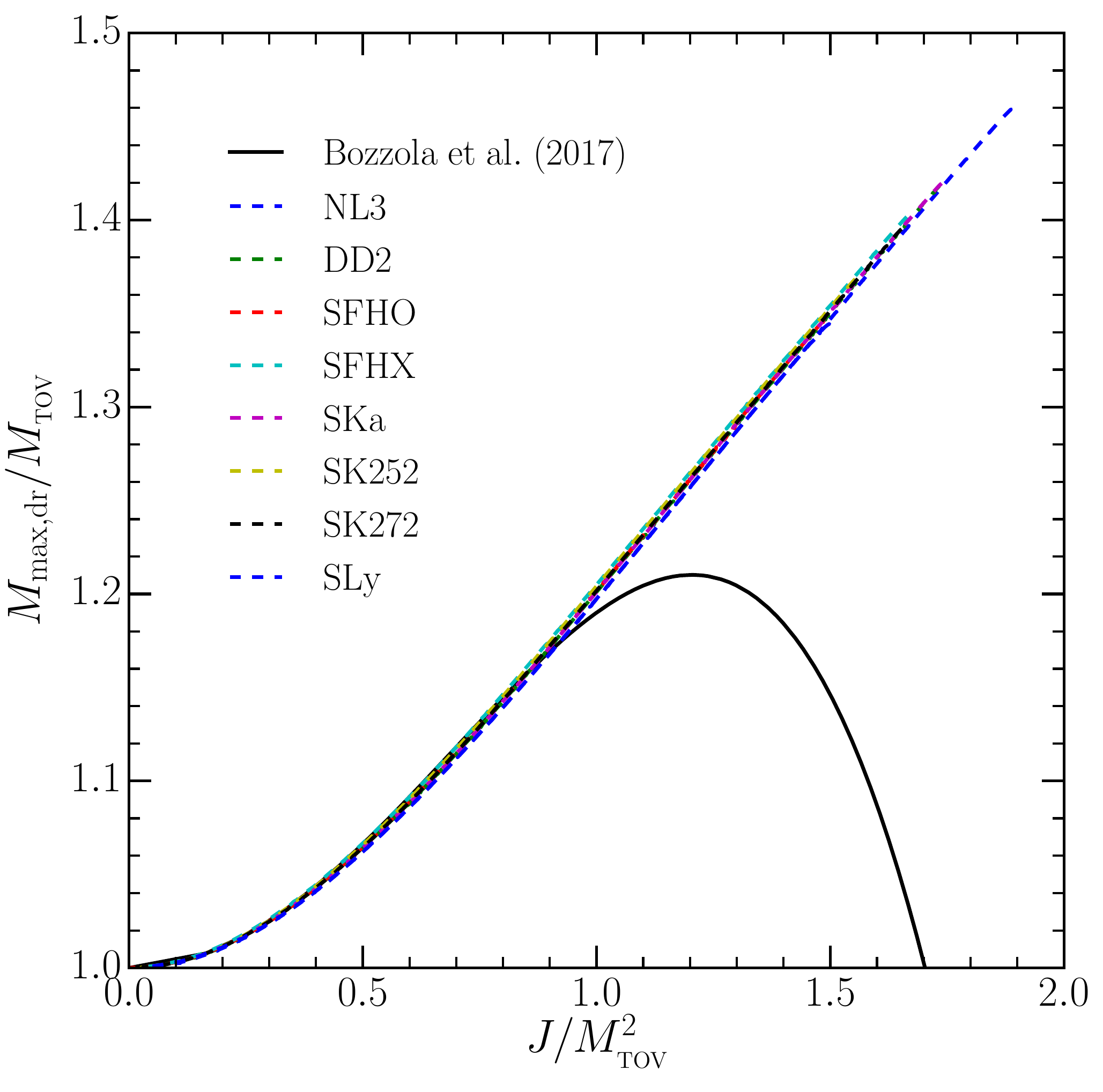}
     \caption{Normalised mass over $J/M_{_{\rm TOV}}^2$. Shown is the
       data for eight EOSs and the comparison with expression (13c) by
       \citet{Bozzola2017}.}
     \label{fig:comparison}
\end{figure}

Before concluding this section we note that a universal relation between
the turning-point mass and the angular momentum was reported recently
also by \citet{Bozzola2017}. The relation proposed in Eq. (13c) of
\citet{Bozzola2017}, however, does not allow to calculate masses larger
than $M_{\rm max, \, dr} \simeq 1.2\,M_{_{\rm TOV}}$, because the data
used for determining the best-fitting function was limited to
$J \in [0,7]$ or, equivalently, $J/M^2_{_{\rm TOV}} \in [0,
  0.7]$. As a result, the fit deviates for neutron stars with higher $J$,
as can be seen from Fig.  \ref{fig:comparison}. Of course, it is possible
to repeat the fit in Eq. (13c) of \citet{Bozzola2017} by increasing the
range for $J$, but this would then depend on the specific EOS, since
different EOSs have different $J_{\rm max}$. Our analysis instead uses 
dimensionless quantities,\ie $j/j_{\rm max}$, and so does not 
introduce arbitrary mass scales.

%

%-----------------------------------------------------------------------
\section{Conclusion}
\label{sec:conc}
%-----------------------------------------------------------------------

We have computed a large number of equilibrium models of differentially
rotating relativistic stars of type A in the classification of
\citet{Ansorg2009}, and evolved selected models along sequences of
constant angular momentum and for two representative degrees of
differential rotation. In this way, we have shown that the
neutral-stability line that marks the onset of dynamical instability for
uniformly rotating neutron stars can be extended also to differentially
rotating ones. In turn, this indicates that all the rotating stars on the
high-density side of the turning point on sequences of constant angular
momentum are unstable solutions. Furthermore, because the
neutral-stability limit is sufficiently close to the turning points, the
turning-point criterion remains a reasonable first approximation to mark
dynamically unstable models, at least for small values of $J$.

Following \citet{Breu2016}, we have also shown that ``quasi-universal''
relations can be found for the turning-point mass, through which we compute
the maximum mass allowed by differential rotation as $M_{\rm max, dr}
\simeq \left(1.54 \pm 0.05\right) M_{_{\rm TOV}}$, with $M_{_{\rm TOV}}$
the maximum nonrotating mass. Finally, we could confirm the validity of
the universal relation derived for the one-parameter $j$-constant law by
\citet{Bozzola2017}, but also point out that such validity holds true
only for a limited range of $J$.

As a final remark, we note that the analysis carried out here adopts the
simple and commonly used law of differential rotation
\eqref{eq:rotlaw}. However, binary neutron-star merger simulations have
shown that the merged object has a rather different rotation profile,
with a maximum off the centre [see, \eg \citet{Kastaun2014,
    Hanauske2016}]. Recently, a new rotation law has been proposed that
yields a rotation profile similar to the one from merger simulations
\citep{Uryu2017}. The constant angular-momentum sequences calculated with
this new law also exhibit turning points, suggesting that our results
will hold qualitatively also with more realistic rotation profiles.

\section*{Acknowledgements}
We thank G. Bozzola and C. Breu for useful discussions. This research is
supported in part by the ERC synergy grant ``BlackHoleCam: Imaging the
Event Horizon of Black Holes" (Grant No. 610058), by ``NewCompStar'',
COST Action MP1304, by the LOEWE-Program in the Helmholtz International
Center (HIC) for FAIR, by the European Union's Horizon 2020 Research and
Innovation Programme (Grant 671698) (call FETHPC-1-2014, project
ExaHyPE). The simulations were performed on the SuperMUC cluster at the
LRZ in Garching, on the LOEWE cluster in CSC in Frankfurt, on the
HazelHen cluster at the HLRS in Stuttgart.

\bibliographystyle{mnras}
\bibliography{aeireferences}

%\label{lastpage}
\end{document}